\documentclass[showpacs,amsmath,amssymb,twocolumn,pra,superscriptaddress,nofootinbib]{revtex4-1}

\usepackage[dvips]{graphicx} 
\usepackage{amsfonts}
\usepackage{amssymb}
\usepackage{amscd}
\usepackage{amsmath}    
\usepackage{enumerate}
\usepackage{epsfig}
\usepackage{subfigure}
\usepackage{xcolor}
\usepackage{amsthm}
\usepackage{multirow}
\usepackage{graphicx}  
\usepackage{epstopdf}  
\usepackage{graphicx}
\usepackage{dcolumn}
\usepackage{bm}
\usepackage{lipsum}
\usepackage{tabularx}
\usepackage{array}
\usepackage{mathrsfs}
\usepackage{amssymb}
\usepackage{amsmath}
\usepackage{cases}  
\usepackage{graphicx}
\usepackage{subfigure} 
\usepackage{dcolumn}
\usepackage{bm}
\usepackage{verbatim} 
\usepackage{graphicx}  
\usepackage{epstopdf}  

\usepackage{hyperref}

\begin{document}

\title{Robust countermeasure against  detector control attack in practical quantum key distribution system
 }

\author{Yong-Jun Qian}
\address{CAS Key Laboratory  of Quantum Information,  University of Science and Technology of China, Hefei  230026, China}
\address{CAS Center for Excellence in Quantum Information and Quantum Physics, University of Science  and Technology of China, Hefei 230026,  China}
\address{State Key Laboratory of Cryptology, P. O. Box 5159, Beijing 100878,  China}
\address{These authors contributed equally to this work}
\author{De-Yong He}
\address{CAS Key Laboratory  of Quantum Information,  University of Science and Technology of China, Hefei  230026, China}
\address{CAS Center for Excellence in Quantum Information and Quantum Physics, University of Science  and Technology of China, Hefei 230026,  China}
\address{State Key Laboratory of Cryptology, P. O. Box 5159, Beijing 100878,  China}
\address{These authors contributed equally to this work}
\author{Shuang Wang}
\email{wshuang@ustc.edu.cn }
\address{CAS Key Laboratory  of Quantum Information,  University of Science and Technology of China, Hefei  230026, China}
\address{CAS Center for Excellence in Quantum Information and Quantum Physics, University of Science  and Technology of China, Hefei 230026,  China}
\address{State Key Laboratory of Cryptology, P. O. Box 5159, Beijing 100878,  China}
\author{Wei Chen}
\address{CAS Key Laboratory  of Quantum Information,  University of Science and Technology of China, Hefei  230026, China}
\address{CAS Center for Excellence in Quantum Information and Quantum Physics, University of Science  and Technology of China, Hefei 230026,  China}
\address{State Key Laboratory of Cryptology, P. O. Box 5159, Beijing 100878,  China}
\author{Zhen-Qiang Yin}
\address{CAS Key Laboratory  of Quantum Information,  University of Science and Technology of China, Hefei  230026, China}
\address{CAS Center for Excellence in Quantum Information and Quantum Physics, University of Science  and Technology of China, Hefei 230026,  China}
\address{State Key Laboratory of Cryptology, P. O. Box 5159, Beijing 100878,  China}
\author{Guang-Can Guo}
\address{CAS Key Laboratory  of Quantum Information,  University of Science and Technology of China, Hefei  230026, China}
\address{CAS Center for Excellence in Quantum Information and Quantum Physics, University of Science  and Technology of China, Hefei 230026,  China}
\address{State Key Laboratory of Cryptology, P. O. Box 5159, Beijing 100878,  China}
\author{Zheng-Fu Han}
\address{CAS Key Laboratory  of Quantum Information,  University of Science and Technology of China, Hefei  230026, China}
\address{CAS Center for Excellence in Quantum Information and Quantum Physics, University of Science  and Technology of China, Hefei 230026,  China}
\address{State Key Laboratory of Cryptology, P. O. Box 5159, Beijing 100878,  China}

\begin{abstract}
In real-life implementations of quantum key distribution (QKD), the physical systems with unwanted imperfections would be exploited by an eavesdropper. Based on imperfections in the detectors, detector control attacks have been successfully launched on several QKD systems, and attracted widespread concerns. Here, we propose a robust countermeasure against these attacks just by introducing a variable attenuator in front of the detector. This countermeasure is not only effective against the attacks with blinding light, but also robust against the attacks without blinding light which are more concealed and threatening. Different from previous technical improvements, the single photon detector in our countermeasure model is treated as a blackbox, and the eavesdropper can be detected by statistics of the detection and error rates of the QKD system. Besides theoretical proof, the countermeasure is also supported by an experimental demonstration. Our countermeasure is general in sense that it is independent of the technical details of the detector, and can be easily applied to the existing QKD systems.
\end{abstract}


\maketitle

\section{Introduction}
\label{Introduction}

Quantum key distribution (QKD) is one of the most promising applications of quantum information. It enables two parties, Alice and Bob, to exchange key bits, whose security  has been theoretically proved \cite{Quantumcryptography2002,lo1999unconditional,RevModPhys2009,
QuantumInfComput2004,Advances2019}. Unfortunately, real-life QKD systems still can be hacked due to the imperfection of the devices. Several kinds of attacks have been reported, such as photon-number-splitting (PNS) attack \cite{PNS1995,PNS2002}, phase-remapping attack \cite{phase-remapping2007,phase-remapping2010}, Trojan-horse attack \cite{gisin2006trojan}, time-shift attack \cite{time-shift2007,time-shift2008}, wavelength-dependent attack \cite{wavelength-dependent2011} and detector control attack \cite{NatPhotonics2010,NatureCommun2011,thermalblind2010,after-gate2011,faint2011,laserdamage2014,passivequench2009,activequench2011,SNSPD2011,self-differencing2013,ATR}. In order to defense these attacks, many countermeasures are  also proposed, some of which  are very effective and can even perfectly defense the corresponding attack, for example, decoy-state method is a perfect solution    against PNS attack \cite{decoy,decoy1,decoy2,decoy3}.

Recently, most attacks  focus on the measurement equipment, especially on single-photon detectors (SPDs). Among which detector control attack is  the most fatal one and has attracted lots of attentions \cite{NatPhotonics2010,NatureCommun2011,thermalblind2010,after-gate2011,faint2011,laserdamage2014,passivequench2009,activequench2011,SNSPD2011,self-differencing2013,ATR}.  To implement a detector control attack, Eve randomly chooses  bases to measure the quantum states sent from Alice, then  resends  the results using trigger light with specific optical power. Due to the control effect of trigger pulses,  the outputs of Bob's detectors are nearly identical to Eve's. This will cause zero or little extra error bits,  and Eve can obtain a copy of  raw keys without being revealed by  legitimate users.  Note that not all detector-related attacks belong to the detector control attack, such as the detector dead time attack \cite{weier2011quantum} and time-shift attack \cite{time-shift2007,time-shift2008} are not within the scope of the detector control attack.  In one of the most typical experiments \cite{NatPhotonics2010}, Eve first uses bright continuous-wave  illumination to blind the SPDs, and converts them into  linear detectors which are not sensitive to single photon. Then Eve  can  fully control the SPDs by sending trigger pulses  that superimposed with the blinding light. There are a series of experiments using  both blinding light and trigger light, such as continuous-wave blinding attack \cite{NatPhotonics2010,NatureCommun2011}, thermal blinding attack \cite{thermalblind2010,activequench2011}, sinkhole blinding attack \cite{thermalblind2010}. Furthermore, Eve need only send trigger light pulses to directly control SPDs, such as after-gate attack \cite{after-gate2011}, faint-after-gate attack \cite{faint2011}, detector control attack under specific  laser damage \cite{laserdamage2014}. These detector control attacks without blinding light   are more concealed and threatening to QKD systems  than the ones with blinding light. In this sense, more attentions should be paid to  detector control attack without blinding light.

There are various countermeasures to defense the detector control attack. The most ideal one is device-independent scheme \cite{acin2007device,masanes2011secure}, which excludes all the imperfections of the devices, but still impractical to applications of real-world use under current techniques.  The most effective one is measurement-device-independent scheme \cite{MDI2012,Pirandola2012}, but it requires a Bell-state measurement of two independent remote laser sources, which is experimentally challenging.  The other methods  are mainly focusing on the technical improvements in SPDs \cite{lydersen2011secure,Randomvariation2015,elezov2015countermeasures,lee2016countermeasure} or measurement devices \cite{honjo2013countermeasure,wang2016countermeasures,da2015safeguarding,maroy2016secure}, or passively monitoring  parameters \cite{yuan2011resilience,lydersen2011comment,yuan2011response,Real-timemonitoring2012,SWang2014,fujiwara2013characteristics,meda2016backflash}.  However, these countermeasures may not be provably secure because the characteristics of actual devices and implementations are not under consideration in the security proofs \cite{lydersen2011comment,lo2014secure}. Furthermore, some countermeasures may be available to a kind of SPDs \citep{Koehler2018} or effective to one specific attack, but not all types of detector control attack. For example, the method of monitoring the photocurrent of the avalanche photodiode  is effective to find the detector control attack with blinding light, but will fail to detect the recent avalanche-transition region attack \cite{ATR}.  Another lately  proposed countermeasure is to randomly remove gates and check the clicks in the absence of the gates \cite{Randomvariation2015}, while Eve can still implement traceless control of SPDs \cite{huang2016testing}, since the method  causes changes of both the gate signal leakage and gain factor in SPD circuits.

To defense detector control attack, we propose a robust countermeasure model by introducing a variable  attenuator (VA) in front of the SPD. With the random change of attenuation of VA and the analysis of the corresponding detection events and errors, the countermeasure criteria is proven effective against the detector control attack without blinding light.    An experiment is also demonstrated to support the effectiveness of the VA-SPD countermeasure.   If Eve implements the detector control attack with blinding light, she would introduce new fingerprints in addition to high photocurrent, and   trigger the alarm of the QKD system.

\section{countermeasure model}
\label{model}

\begin{figure}[hbt]
\centering
\includegraphics[width=8.5cm]{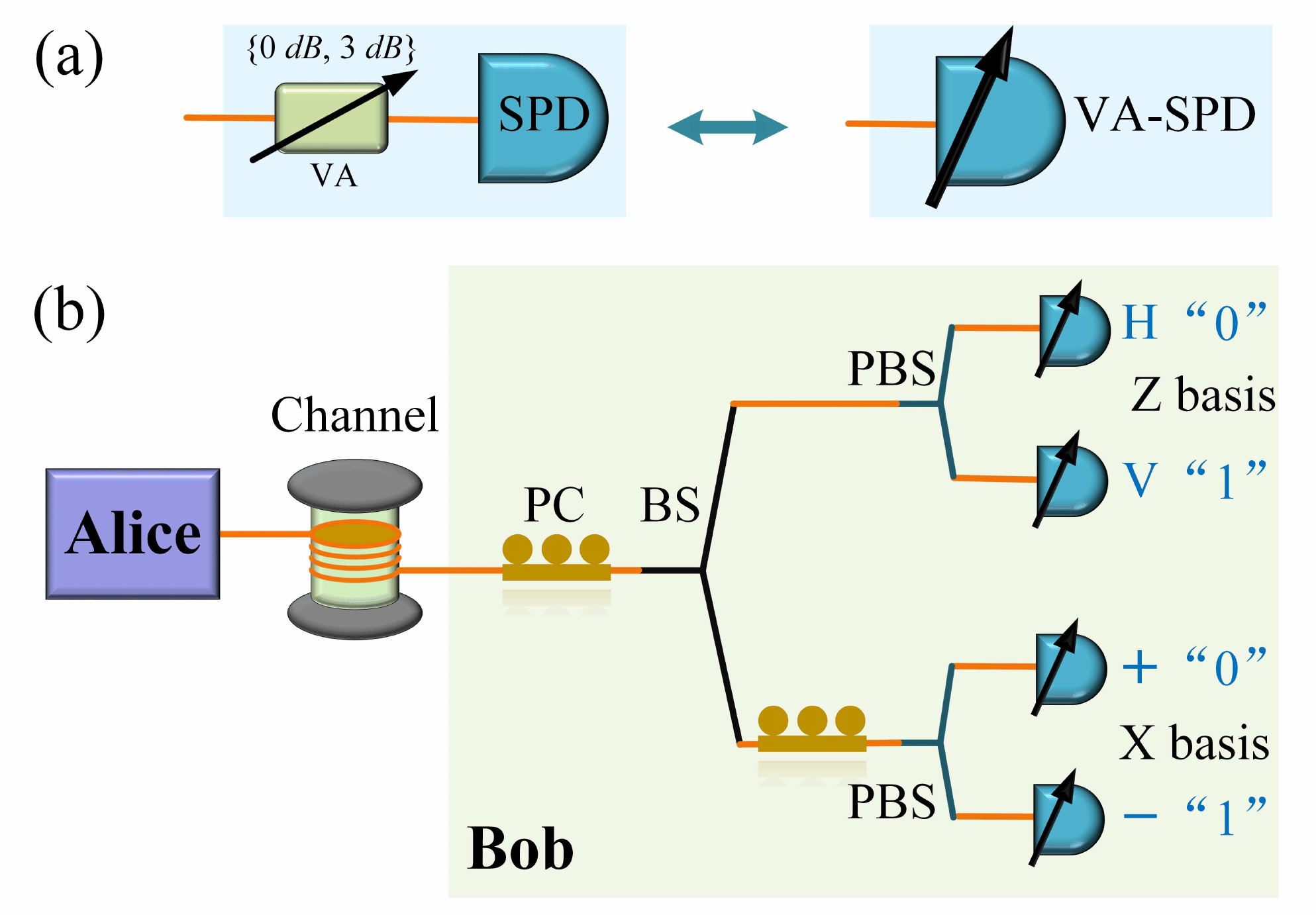}
\caption{\label{fig:model}(a) Schematic of the countermeasure model. The VA-SPD model consists of a variable attenuator(VA) and a single photon detector(SPD). (b) Application of the VA-SPD model to a  polarization-encoding BB84 system. PC, polarization controller; BS, beamsplitter; PBS, polarization beamsplitter.}
\label{Fig:model}
\end{figure}

The implementation procedure of our countermeasure model is shown in Fig. \ref{fig:model}(a), a VA is placed in front of the SPD, and its attenuation can be randomly changed among several values. Note that the number of attenuation values is at least two, and the value difference is $3\ dB$ (see APPENDIX A for an explanation of the necessity of $3\ dB$). In this paper, the number of attenuation values is two, $0\ dB$ and $3\ dB$, respectively. The countermeasure model is named as VA-SPD, which is suitable for different kinds of SPDs, such as photomultiplier tubes (PMT), superconducting single-photon detector (SSPD) and  semiconductor detectors (Si SPD, InGaAs/InP SPD). In the  model, SPD is treated as a blackbox, which only has two ports: an optical signal input and a detection output. It is not necessary to modify the internal circuits or monitor the technical parameters in the SPD. So our countermeasure model is applicable for prepare-and-measure  QKD systems, just  by directly replacing the original SPD with a VA-SPD.

In order to explicitly illustrate the procedure of the proposed countermeasure model, we apply the VA-SPD to  a typical polarization-encoding BB84  system with  passive measurement bases selection, which  has been hacked by several attacks \cite{wavelength-dependent2011,NatureCommun2011,liu2014universal,lamas2007breaking,nauerth2009information,sajeed2015security,weier2011quantum}. As  shown in Fig. \ref{fig:model}(b),  Alice prepares and sends Bob a sequence of polarization states, each randomly chosen from four polarization states \{$H$, $V$, $+ $, $ - $\}, $H$ and $V$ are horizontal and vertical polarization states, respectively.  $+$ and $-$ denote $+45^\circ$ and $-45^\circ$ linear polarization states, respectively. For each state, Bob passively and randomly chooses one of the two measurement bases \textemdash $ $  Z (or rectilinear) basis and  X (or diagonal) basis \textemdash $ $  to project the input photon into one of these four polarization states.   At Bob's site,  each VA-SPD corresponds to one polarization state, and the attenuation value of VA in each VA-SPD is randomly set to  $0\ dB$ or $3\ dB$.   After the announcement of basis choices, we can  get the detection rate and quantum bit error rate (QBER) for each detector. Different from the original system, two kinds of results could be obtained for two values of VA in the VA-SPD. Here, for each VA-SPD,  \{$R_{0}$, $R_{3} $\} and \{$e_{0}$, $e_{3}$\} denote the detection rates and QBERs with $0\ dB$ and $3\ dB$ attenuation value, respectively. 

 In  almost all BB84 QKD systems,  weak coherent states sources  are widely used. The    photon number of each pulse prepared by Alice follows a  Poisson distribution\cite{decoy3}. Suppose the expected photon number of each pulse is $\mu$, the overall transmission and detection efficiency   between Alice and Bob is $\eta$, and the background rate is $Y_{0}$, then  the detection rate is given by
 \begin{equation}
 R=1-(1-Y_{0})e^{-\mu\eta},
 \label{eq:Rno}
\end{equation}
which means the probability that Bob gets one detection count when Alice sends one pulse. For each detector, we can get similar expressions, the only differences are the meaning of $\eta$ and $Y_{0}$. Thus, the ratio between detection rates of one VA-SPD with $0\ dB$ and  $3\ dB$   attenuation certainly satisfies
 
\begin{equation}
\begin{aligned}
1< \alpha \equiv \dfrac{R_{0}}{R_{3}}< 2.
\end{aligned}
\label{eq:R0R3no}
\end{equation}

Additionally,  the  QBERs  with $0\ dB$ and $3\ dB$ attenuation should be less than the threshold to generate secure keys. We have 
\begin{equation}
\begin{aligned}
 \{e_{0}, e_{3}\}< e_{th},
\end{aligned}
\label{eq:e0e3no}
\end{equation}
where $e_{th}$ is the threshold of QBER, and is 11\% for the four-state BB84 system \cite{securityproof1,securityproof2,securityproof3}.

In the BB84 QKD system employed VA-SPDs, the relationships of the  detection rate Eq. \eqref{eq:R0R3no} and QBER  Eq. \eqref{eq:e0e3no} between $0\ dB$ and $3\ dB$ attenuation should be held simultaneously. Here, we prove that the relationships of  Eq. \eqref{eq:R0R3no} and  Eq. \eqref{eq:e0e3no} cannot be satisfied simultaneously if the system was hacked by the detector control attack. This criteria would be a trace to find the detector control attack. And in the VA-SPD countermeasure model, we do not need open the SPD to monitor some specific parameters. Furthermore, the simulation results show that the fingerprint introduced by the detector control attack is pretty obvious.

\subsection{Theoretical proof of the criteria}
In all detector control attacks without blinding light,  Eve first uses a random basis to measure the quantum state sent by Alice, then resends a trigger signal to Bob based on her measurement result. The power of  the trigger signals is not in single-photon level, but in multi-photon level. And the power of the trigger signal stays the same, regardless of Eve's measurement result. If Eve and Bob select matching bases, the trigger signal would hit one detector with full optical power. If Eve and Bob select opposite bases, the trigger signal would be split into two half parts and hit two detectors.   According to the optical power (full, half) hitting the detector and  the attenuation value ($0\ dB$, $3\ dB$) of the VA-SPD,     $P_{f,0} $ is defined as  the detection probability    with full optical power  when the attenuation is $0\ dB$. $P_{f,3} $ is likewise defined  when the attenuation is $3\ dB$; similarly, with half power,  $P_{h,0}$ and $P_{h,3}$ are defined as the detection probabilities when the attenuation are $0\ dB$ and $3\ dB$, respectively. Suppose Eve select two measurement basis with equal probability, the detection rates of Bob's one VA-SPD with $0\ dB$ and $3\ dB$ attenuation are given by

\begin{eqnarray}
\begin{aligned}
R_{0}^{atk}=
\frac{1}{4}P_{f,0}+\frac{1}{2}P_{h,0},
\end{aligned}
\label{eq:R0attack}
\end{eqnarray}

\begin{equation}
\begin{aligned}
R_{3}^{atk}=\frac{1}{4}P_{f,3}+\frac{1}{2}P_{h,3}.
\end{aligned}
\label{eq:R3attack}
\end{equation}
Here, ``atk'' means under the detector control attack.

As an acceptable assumption, both detectors in the same basis are identical here for simplicity. About the QBER of Bob's one VA-SPD, it involves the other or orthogonal detector in the same basis. If both detectors click simultaneously, Bob assigns a random bit value. Since attenuation values of both VA-SPDs are changed independently, there are two circumstances: one is both VA-SPDs have the same attenuation value ($0\ dB$ or $3\ dB$), the other is the attenuation values are opposite ($0\ dB$ \& $3\ dB$, or $3\ dB$ \& $0\ dB$). When both VA-SPDs have the same attenuation value, the QBER of Bob's one VA-SPD with $0\ dB$ and $3\ dB$ attenuation are given by

\begin{equation}
\begin{aligned}
e_{0(s)}^{atk}=\frac{2P_{h,0}-P_{h,0}^{2}}{2P_{f,0}+2(2P_{h,0}-P_{h,0}^{2})},
\end{aligned}
\label{eq:e0attack}
\end{equation}

\begin{equation}
\begin{aligned}
e_{3(s)}^{atk}=\frac{2P_{h,3}-P_{h,3}^{2}}{2P_{f,3}+2(2P_{h,3}-P_{h,3}^{2})}.
\end{aligned}
\label{eq:e3attack}
\end{equation}

For the detector control attack without blinding light, there are two equivalent situations \textemdash $ $ one is Eve and Bob select matching bases and the attenuation value of the corresponding VA-SPD is $3\ dB$, the other is Eve and Bob select opposite bases and the attenuation value of the VA-SPD is $0\ dB$. Then we have $P_{h,0}=P_{f,3}$. From Eqs. \eqref{eq:R0R3no},\eqref{eq:R0attack} and \eqref{eq:R3attack}, we get 
\begin{equation}
\begin{aligned}
P_{f,0}+2P_{h,0}=\alpha(P_{f,3}+2P_{h,3}).
\end{aligned}
\label{eq:R0R3a}
\end{equation}
If the relationship of  Eq. \eqref{eq:e0e3no} is satisfied, we have

\begin{equation}
\begin{aligned}
2P_{h,0}-P_{h,0}^{2}< 2 \alpha e_{th}(P_{h,0}+2P_{h,3})-2e_{th}P_{h,0}^{2},
\end{aligned}
\label{eq:eu0<=eth}
\end{equation}

\begin{equation}
\begin{aligned}
\alpha(2P_{h,3}-P_{h,3}^{2})< 2 \alpha e_{th}(P_{h,0}+2P_{h,3}-P_{h,3}^{2}).
\end{aligned}
\label{eq:eu3<=eth}
\end{equation}
By adding both sides of these inequalities,  we get
\begin{widetext}
\begin{equation}
\begin{aligned}
(2-P_{h,0}-4\alpha e_{th})P_{h,0}+\alpha (2-P_{h,3}-8e_{th})P_{h,3}+2e_{th}P_{h,0}^{2}+2\alpha e_{th}P_{h,3}^{2}< 0.
\end{aligned}
\label{eq:eu0+eu3}
\end{equation}
\end{widetext}
It's obvious that this inequality cannot be satisfied in condition that $0\leq \{P_{h,0}, P_{h,3}\} \leq 1$, $ 1< \alpha < 2$, and $e_{th} < 11\% $. Therefore, it is impossible to satisfy the relationships (Eq. \eqref{eq:R0R3no} and Eq. \eqref{eq:e0e3no}) simultaneously under the detector control attack.

Similarly, when the attenuation values of VA-SPDs are opposite,  the QBER  of Bob's one VA-SPD with $0\ dB$ and $3\ dB$ attenuation are given by

\begin{equation}
\begin{aligned}
e_{0(opp)}^{atk}=\frac{2P_{h,0}-P_{h,0}P_{h,3}}{2P_{f,0}+2(2P_{h,0}-P_{h,0}P_{h,3})},
\end{aligned}
\label{eq:e0attack2}
\end{equation}

\begin{equation}
\begin{aligned}
e_{3(opp)}^{atk}=\frac{2P_{h,3}-P_{h,0}P_{h,3}}{2P_{f,3}+2(2P_{h,3}-P_{h,0}P_{h,3})}.
\end{aligned}
\label{eq:e3attack2}
\end{equation}
We can also prove that the relationships (Eq. \eqref{eq:R0R3no} and Eq. \eqref{eq:e0e3no}) cannot be satisfied simultaneously through a similar process.

\subsection{Simulation results if one relationship is satisfied }
According to the  above proof, the relationships of
Eq. \eqref{eq:R0R3no} and Eq. \eqref{eq:e0e3no} cannot be satisfied simultaneously under the detector control attack. Here, through the approach of numerical simulation, we show that the violation of one relationship would be pretty obvious if the other relationship is satisfied.  Details of the calculation process is in Appendix. B.

\begin{figure}[hbt]
\centering
\includegraphics[width=8cm]{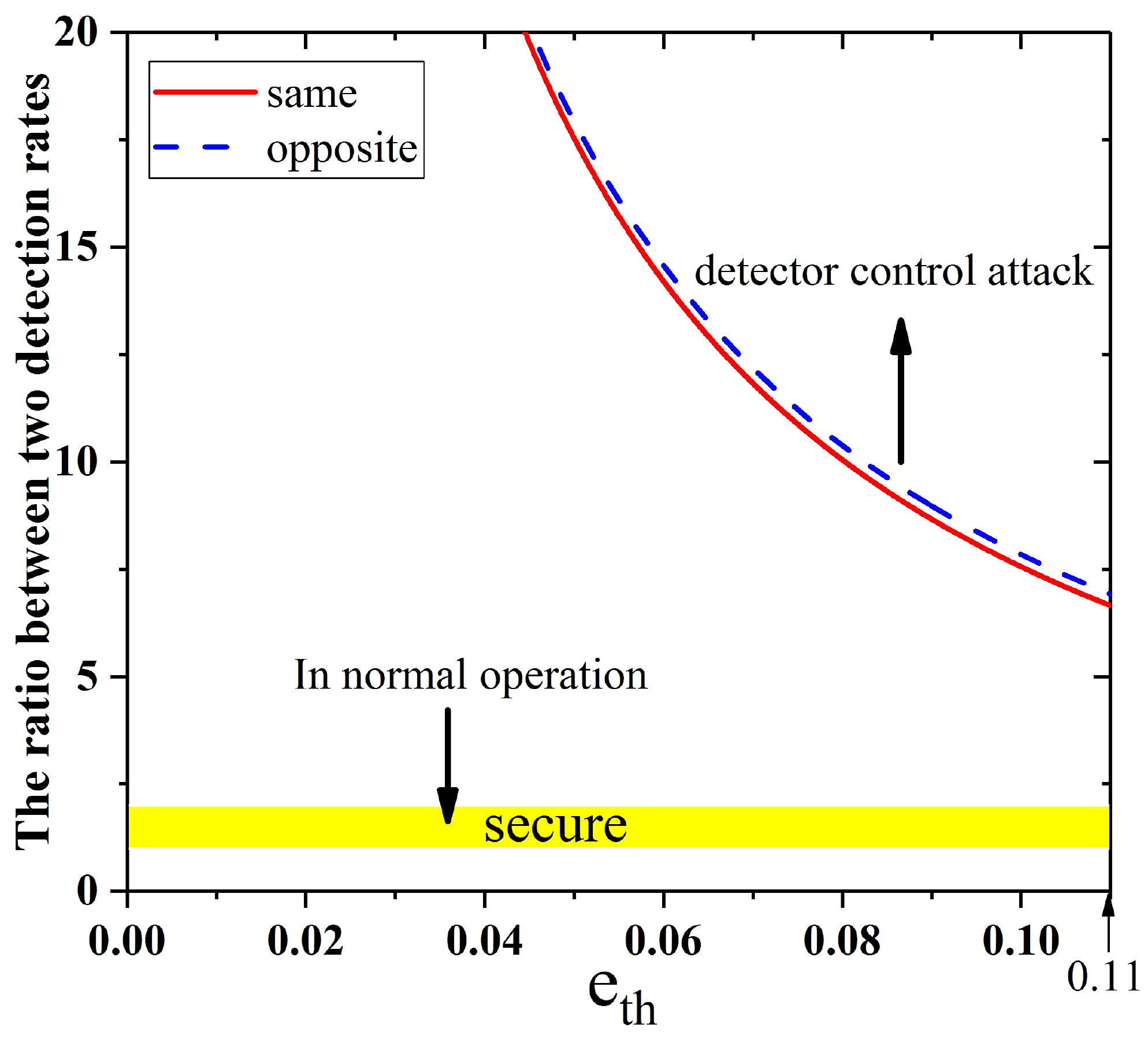}
\caption{Bound of  the ratio between two detection rates when both QBERs are less than $ e_{th} $. }
\label{Fig:euvsratio}
\end{figure}

In the case that the relationship of Eq. \eqref{eq:e0e3no} is satisfied, both QBERs ($e_{0}$ and $e_{3}$) are less than  $e_{th}$. The bounds of the ratio between two detection rates are shown in Fig. \ref{Fig:euvsratio}. For the QKD system in normal operation, the ratio between two detection rates $\alpha=\frac{R_{0}}{R_{3}}$ locates in the yellow region ($1< \alpha < 2$).  While, if the system was under  the detector control attack, the lower bounds of the ratio between two detection rates are depicted by the red line and blue dashed line respectively, the red line ($\frac{R_{0(s)}^{atk}}{R_{3(s)}^{atk}}$) corresponds to the circumstance that both VA-SPDs in the same basis have the same attenuation value (denoted as same), and the blue dashed line ($\frac{R_{0(opp)}^{atk}}{R_{3(opp)}^{atk}}$) corresponds to the situation that these two attenuation values are   opposite (denote as opposite). The slight difference between the red line and the blue dashed line comes from the discrepancy of QBERs in two circumstance (Eqs. \eqref{eq:e0attack}-\eqref{eq:e3attack} and Eqs. \eqref{eq:e0attack2}-\eqref{eq:e3attack2}). Obviously, the ratio between two detection rates under the detector control attack is far from the secure region, and as a good fingerprint, the detector control attack would be detected easily. As the threshold of QBER  $e_{th}$ was set smaller, the ratio between two detection rates  would be greater, and farther from the secure region. Even when $e_{th}$ is 11\%,  the lower bounds of in two situation are more than  6.5, which would be easy to find the attack.

\begin{figure}[hbt]
\centering
\includegraphics[width=8cm]{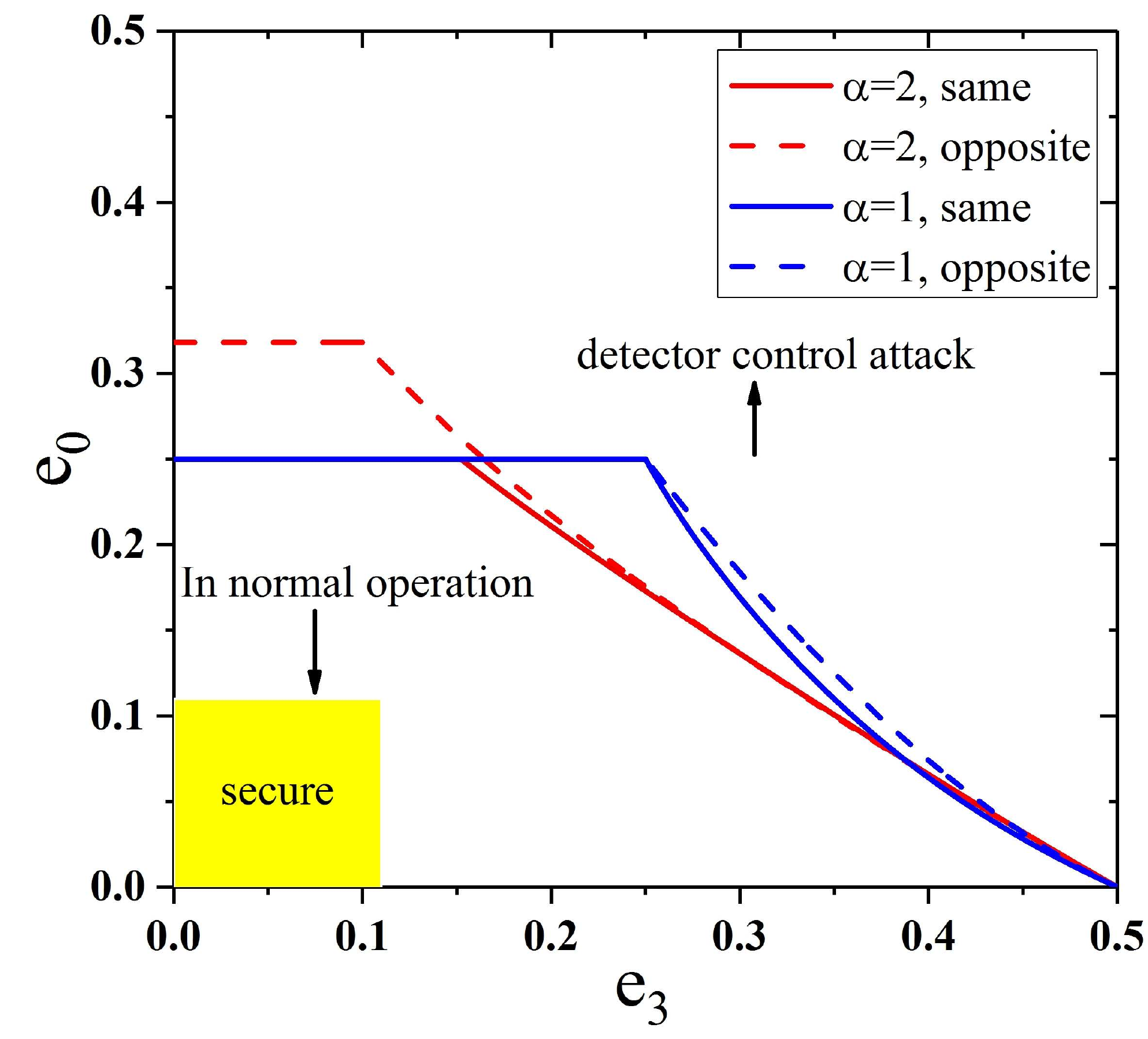}
\caption{Scales of   QBERs with   $0 \ dB$ ($e_{0}$) and $3 \ dB$ ($e_{3}$) attenuation when the ratio between two detection rates $ \alpha $ locates in the secure region.}
\label{Fig:e0vse3}
\end{figure}

In the case that the relationship of Eq. \eqref{eq:R0R3no} is satisfied,  the ratio between two detection rates $\alpha$ locates  in the secure region. Fig. \ref{Fig:e0vse3}  illustrates the scales of  QBERs with  $0 \ dB$ ($e_{0}$) and $3 \ dB$ ($e_{3}$) attenuation. For the system in normal operation, both $e_{0}$ and $e_{3}$ should be less than 11\%, as shown in the yellow region. Under the detector control attack, though  Eve could control transmittance and number of trigger pulses to guarantee the detection rates unchanged, QBERs would increase a lot. In Fig. \ref{Fig:e0vse3},  the  lines and dashed lines correspond to the situation that two attenuation values are the same and opposite, respectively. And, the red ones correspond to $\alpha=2$ (upper bound of the ratio between two detection rates), the blue ones correspond to $\alpha=1$ (lower bound of the ratio). It is obvious that  $e_{0}$ and  $e_{3}$ cannot be in the secure region together. If one of  \{$e_{0}$,  $e_{3}$\} was less than  11\%, the other one would be more than 25\%. Hence, these values of QBER would be very easy to trigger the alarm.

\section{Experimental demonstration of the countermeasure}
\label{application}
To show the effectiveness of the countermeasure, we experimentally apply it against the faint after-gate attack \cite{faint2011}, which is a typical detector control attack without blinding light. The experimental setup is similar to the schematic depiction in Fig. \ref{Fig:model}(b). In the normal operation, Alice is the sender, who sends a sequence of polarization states with a repetition rate of 5 MHz and a expected photon number $\mu=0.1$ of each pulse. At Bob's site, the random attenuation values of VA in VA-SPD are  $0\ dB$ and $3\ dB$, and the insertion loss  of VA is approximately $0.6\ dB$, which reduces the original detection efficiency of SPD 12.6\% to a  equivalent detection efficiency of VA-SPD 11.0\% at $0\ dB$ attenuation. To simplify the experiment, Bob only monitors the detection events in Z basis $\{H,V\}$, and always sets the same attenuation value ($0\ dB$ or $3\ dB$) in the corresponding two VA-SPDs (the VA-SPD of bit ``0'' is used to detect the $H$ state, the VA-SPD of bit ``1'' is used to detect the $V$ state.) Hence, for the QKD system in the normal operation, the ratio between two detection rates $\alpha=\frac{R_{0}}{R_{3}}$ is approximately 1.994, and QBERs with $0\ dB$ and $3\ dB$ attenuation values are 1.82\% and 1.91\%, respectively. While, under the faint after-gate attack, Eve becomes the sender, here we skip the intercept and measurement process for simplicity. Different from Alice, Eve needs first measure the characteristic of each SPD, and then carefully control the delay and incident flux of her encoded pulses, the delay makes these pulses arrive after the gate, and the incident flux (a few hundreds photons per pulse) offers superlinearity of the detection probability with full and half optical powers. 

 In order to keep the QBERs below the threshold,  Eve   chooses  the  attack positions  at the falling edge of 0.74 ns and 0.88 ns for the VA-SPDs of bit ``0''  and bit  ``1'', respectively, and the incident flux of 108 photons per pulse. After measuring Eve's encoded pulses in Z basis, the detection probabilities of Bob's two VA-SPDs are  listed in TABLE~\ref{tab:table1}. Taking the VA-SPD of bit ``0'' for example, the detection probabilities with full and half incident flux are  $P_{f,0}=0.10675$ and $P_{h,0}=0.0142$, respectively, when the attenuation value is $0 \ dB$, and are $P_{f,3}=0.01415$ and $P_{h,3}=0.00182$  when the attenuation value is  $3\ dB$. Now the QBERs with two attenuation values are   $e_{0(s)}^{atk}=10.45\%$ and $e_{3(s)}^{atk}=10.22\%$, respectively. Both QBERs are below the threshold 11\%, but the ratio between two detection rates is $\frac{R_{0(s)}^{atk}}{R_{3(s)}^{atk}}=7.60$, far beyond the value in the normal operation. Thus this large ratio between two detection rates would make Bob detect the faint after-gate attack easily.

\begin{table*}[hbt]
\caption{\label{tab:table1} Detection probabilities of Bob's two VA-SPDs when Eve makes both QBERs below the threshold.   (a) VA-SPD of bit ``0'' with two attenuation values. (b) VA-SPD of bit ``1'' with two attenuation values. The first column represents  the polarization state  sent by Eve. }
\begin{ruledtabular}
\begin{tabular}{ccccc}
                                       &\multicolumn{2}{c}{(a) VA-SPD of bit ``0''}        &\multicolumn{2}{c}{(b) VA-SPD of bit ``1''}\\
 Eve$\rightarrow$       &    $0 \ dB $    & $3 \ dB $  & $0 \ dB $    &  $3 \ dB $\\ \hline
 $H$               &  {0.10675}            &    {0.01415}        &  {0.00016}          &{0.00011}  \\
 $+$              &    {0.01413}             &   {0.00181}            &{0.0142} &{0.0018} \\
 $V$              & {0.0002}                &{0.00019}               &{0.10819}  &{0.0144}\\
 $-$              &{0.01427}                 &{0.00183}               &{0.01416}  &{0.00178}\\
\end{tabular}
\end{ruledtabular}
\end{table*}

The other attacking strategy is to keep the detection rates around the values in normal operation. This time Eve chooses the attack positions at the falling edge of 0.68 ns and 0.84 ns for the VA-SPDs of bit ``0''  and bit  ``1'', respectively, and the incident flux of 300 photons per pulse. After measuring Eve's encoded pulses in Z basis, the detection probabilities of Bob's two VA-SPDs are listed in    TABLE~\ref{tab:table2}. Still taking the VA-SPD of bit ``0''  for example,  $P_{f,0}=0.9999$, $P_{h,0}=0.5016$, and $P_{f,3}=0.5015$, $P_{h,3}=0.2421$.    Now the ratio between two detection rates is $\frac{R_{0(s)}^{atk}}{R_{3(s)}^{atk}}=2.03$, close to the value in normal operation. But the QBERs with two attenuation values are   $e_{0(s)}^{atk}=21.46\%$, $e_{3(s)}^{atk}=22.95\%$, both  QBERs are more than ten times of the ones in normal operation. Hence this high QBERs would be pretty obvious to find the trace of Eve.

\begin{table*}[hbt]
\caption{\label{tab:table2} Detection probabilities of Bob's two VA-SPDs when Eve makes the detection rates around the values in normal operation. (a) VA-SPD of bit ``0'' with two attenuation values. (b) VA-SPD of bit ``1'' with two attenuation values. The first column represents  the polarization state sent by Eve.}
\begin{ruledtabular}
\begin{tabular}{ccccc}
                                       &\multicolumn{2}{c}{(a) VA-SPD of  bit ``0'' }        &\multicolumn{2}{c}{(b) VA-SPD of  bit ``1''}\\
 Eve$\rightarrow$        &    $0 \ dB $    & $3 \ dB $  & $0 \ dB $    &  $3 \ dB $\\ \hline
 $H$               &  {0.9999}            &    {0.5015}        &  {0.0009}          &{0.0004}  \\
 $+$              &    {0.5019}             &   {0.2417}            &{0.5018} &{0.2413} \\
 $V$              & {0.0011}                &{0.0005}               &{0.9999}  &{0.5024}\\
 $-$              &{0.5012}                 &{0.2425}               &{0.5023}  &{0.2415}\\
\end{tabular}
\end{ruledtabular}
\end{table*}

Compared with normal QKD systems, the insertion loss of VA and the setting attenuation value would introduce extra attenuation and reduce the key rate. Nevertheless, these impacts can be weakened by choosing proper device and controlling the probability of setting attenuation. About the insertion loss of VA, there is no need to use high-speed intensity modulator since VA is controlled by Bob. In the experiment, the insertion loss of VA is only $0.6\ dB$. About the impact of setting attenuation, the probability of setting $3\ dB$ could be very low in practice, and we can also reduce the impact of statistical fluctuation by accumulating longer time.

\section{Effectiveness against the attack with blinding light}
\label{fingerprint}
In the proof of the criteria of our countermeasure, there is an assumption that $P_{h,0} = P_{f,3}$, which holds in the detector control attack without blinding light, but might fail in the attacks with blinding light. Hence the criteria of our countermeasure might not be deduced. However, in addition to high photocurrent \cite{yuan2011response}, a new fingerprint would be introduced by the attack with blinding light in our countermeasure. So the proposed countermeasure is still effective against the attack with blinding light. 

When the continuous-wave (CW) blinding light enters the VA-SPD, it is first modulated by the VA into full power or half power as the attenuation value is randomly set to $0\ dB$ or $3\ dB$. In order to always blind the SPD, the modulated half power should be above the blinding power of the SPD, which is typically about dozens of microwatt  \cite{huang2016testing}. After the modulated blinding light hits the avalanche photodiode (APD), it will create a modulated train of photocurrent. Every time when the attenuation value of VA changes, a negative ($3\ dB$ $ \rightarrow $ $0\ dB$) or positive ($0\ dB$ $ \rightarrow  $ $3\ dB$) signal will be generated at the output of  the APD. This is the fingerprint left by the blinding light in the VA-SPD. And due to the relatively strong optical power of the blinding light, this fingerprint is fairly obvious and easy to be detected. Furthermore, superimposed with the capacitive noise of the gated APD, this fingerprint would exceed the discrimination voltage, and produce one click (or a few clicks), which has 50\% probability to generate an error bit. Thus, even if we do not monitor the negative or positive signal at the output of the APD, the additional clicks introduced by the blinding light would increase the QBER and yield of the QKD system.

  \begin{figure}[hbt]
\centering
\includegraphics[width=8.5cm]{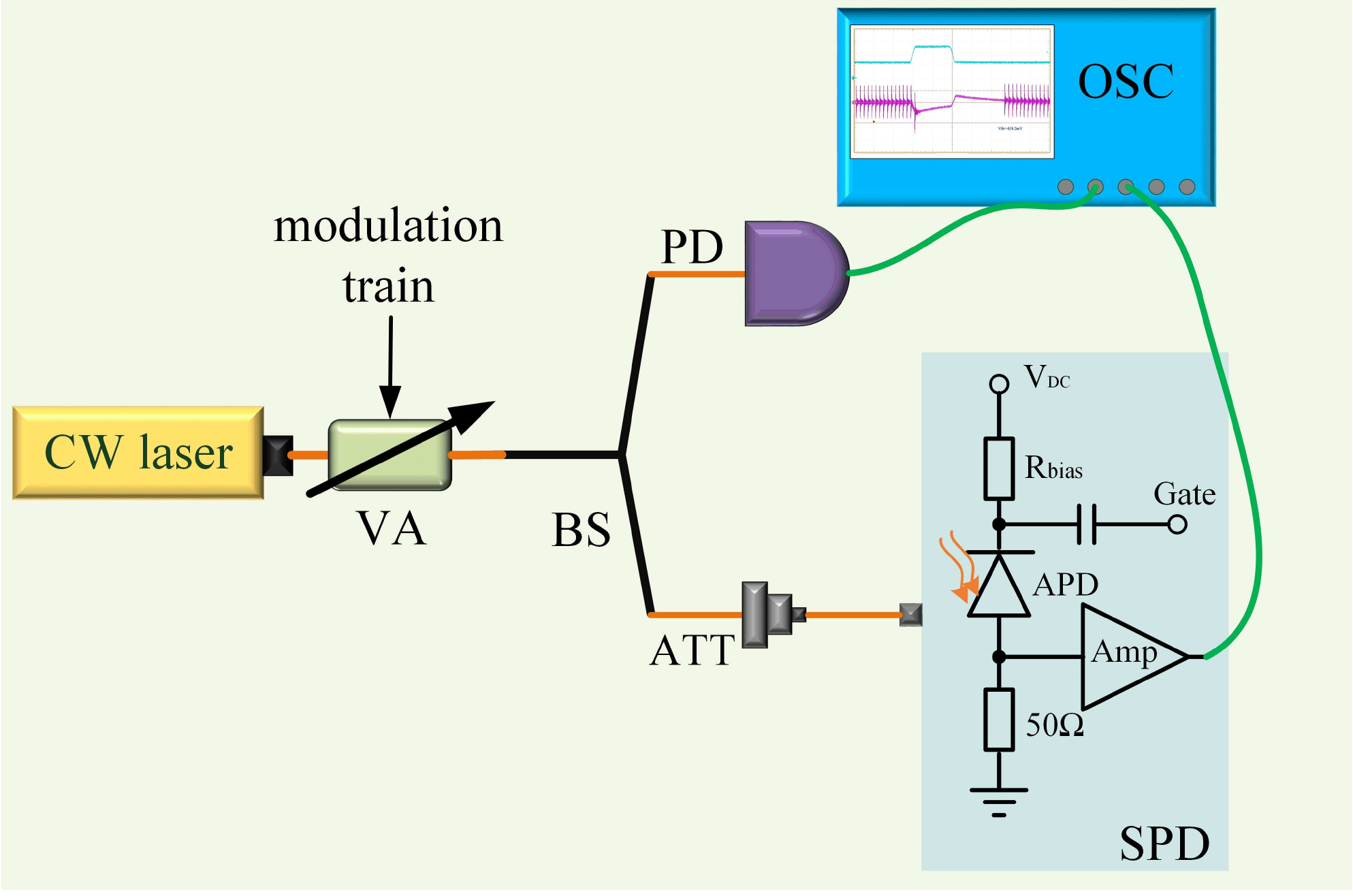}
\caption{ Experimental demonstration of the effectiveness  of the VA-SPD model against the attack with blinding light. VA, variable attenuator; BS, beamsplitter; PD: high-speed photodiode; ATT, optical attenuator; Amp, amplifier;  APD, avalanche photodiode;  OSC, oscilloscope. }
\label{Fig:confirmatoryexperiment}
\end{figure}

\begin{figure}[hbt]
\centering
\includegraphics[width=7.5cm]{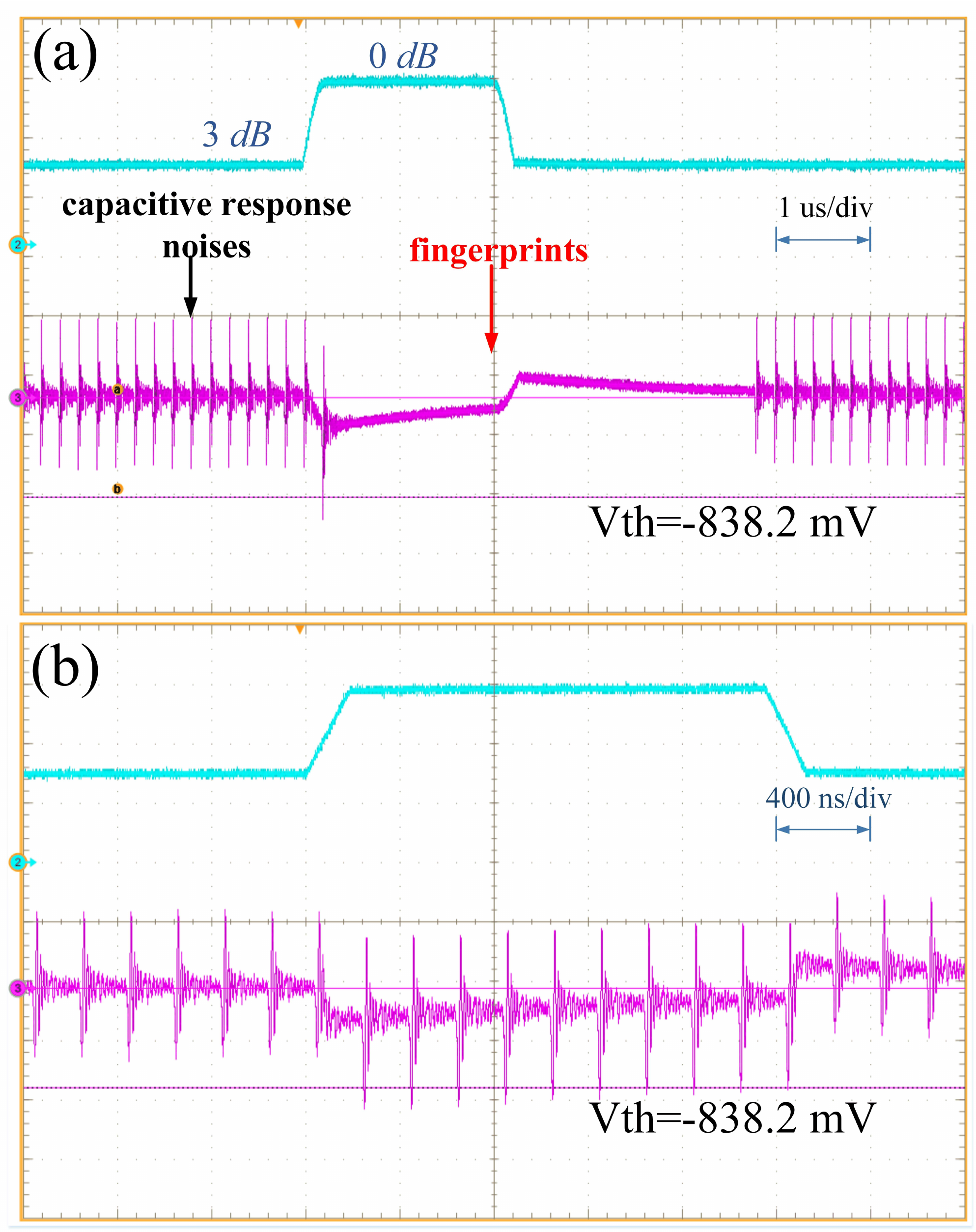}
\caption{ Results recorded by the oscilloscope.  (a) The deadtime of SPD is  4.5 $\mu$s  . (b) The deadtime  is set to bypass. The blue lines refer to the modulated blinding light, and the fuchsine ones refer to the electrical signals  at the output of the APD.  }
\label{Fig:result}
\end{figure}

Although the detector is treated as a blackbox in the countermeasure model, as shown in  Fig. \ref{Fig:confirmatoryexperiment}, we open a detector and measure the output voltage to experimentally demonstrate the reason why Eve's attack with blinding light would also be found.  The CW laser modulated by the VA is splitted into two parts, one part enters a high-speed photodiode (PD) to show the characteristic of the modulated blinding light, the other part is first attenuated by an optical attenuator (ATT) to proper power, and then enters the APD, whose corresponding electrical signals are recorded by an oscilloscope to show the characteristic of the fingerprint left by the blinding light. The attenuation value of the VA is set to  $0\ dB$ or $3\ dB$, and  the response time of VA is approximately $120$ ns. The  SPD is operated at a  frequency of 5 MHz, and  can be  blinded by a CW light at 1550 nm with a power from  11 $\mu$W to 50 $\mu$W. The results observed by the oscilloscope are shown in Fig. \ref{Fig:result}, the blue lines correspond to the modulated blinding light, and the fuchsine ones correspond to the electrical signals  at the output of the APD. Here, the modulated blinding light that enters the APD has the power of 30 $\mu$W (15 $\mu$W) when the attenuation value of VA is $0\ dB$ ($3\ dB$). In Fig. \ref{Fig:result}(a), the deadtime of SPD is 4.5 $\mu$s, so we can observe the fingerprint left by the modulated blinding light clearly. Outside the modulated window, the blinding light is CW, there are only capacitive noises corresponding to the gating pulses. When the intensity changes from 15 $\mu$W to 30 $\mu$W, a negative signal is generated, and then decays during the unchanged intensity; when the intensity changes from 30 $\mu$W to 15 $\mu$W, a positive  signal is generated, and also then decays during the unchanged intensity. Superimposed with the capacitive noises, the negative signal   exceeds the discrimination voltage of -838.2 $mV$,  and produces one click. In Fig. \ref{Fig:result}(b), the deadtime of SPD is set to bypass,  the negative part superimposed with the capacitive noises produce about  9 clicks (fuchsine curve). These abnormal clicks  can be easily detected by the VA-SPD, and trigger the alarm of the QKD system.

\section{conclusion}
\label{conclusion}
In this paper, we have proposed an effective countermeasure  against the  detector control attacks.  After introducing a VA in front of the SPD, the VA-SPD model can detect the detector control attacks easily through analysis of the detection rates and QBERs corresponding to different attenuation values. We first focus on the detector control attacks without blinding light, which are more concealed and threatening to QKD systems. The criteria is proved that the relationships of the detection rate and QBER between $0\ dB$ and $3\ dB$ attenuation cannot be satisfied simultaneously once the system is hacked. In this countermeasure model against the detector control attack, the SPD is treated as a blackbox, we don't need open it to monitor some specific parameters. By numerical simulations and the experimental application against the faint after-gate attack, we not only demonstrate the effectiveness of the VA-SPD model, but also show the obviousness of the fingerprint introduced by Eve. Furthermore, for the detector control attack with blinding light, we analyse and experimentally test the effectiveness of the VA-SPD model, in which a new fingerprint would be introduced in addition to high photocurrent. The countermeasure can be easily applied to the existing QKD system, and would provide a perfect balance between security and practicality.
\section*{Acknowledgments}
This work has been supported by the National Key Research and Development Program of China (Grant Nos.2018YFA0306400), the National Natural Science Foundation of China (Grant Nos. 61622506, 61575183, 61822115, 61775207, 61627820), Anhui Initiative in Quantum Information Technologies.

Y-J. Q. and D-Y. H. contributed equally to this work.

\section*{Appendix A: analysis of the difference of VA's value}
In this section, we explain the reason why the VA's value should differ $3\ dB$.   Assume that the  attenuation value of VA in each VA-SPD is randomly set to  $x\ dB$ and $y\ dB$ ($x<y$). As mentioned above in Sec. \ref{model}, after the announcement of basis choices, \{$R_{x}$, $R_{y} $\}  and \{$e_{x}$, $e_{y} $\}  denotes the detection rates and QBERs with $x\ dB$ and $y\ dB$, respectively. For the QKD system in normal operation, the ratio between detection rates of one VA-SPD with $x\ dB$ and $y\ dB$ attenuation satisfies

\begin{equation}
\begin{aligned}
\alpha^{\ast}=\dfrac{R_{x}}{R_{y}}>1,
\end{aligned}
\label{eq:Rx,mnoRy,mno}
\end{equation}
similarly, the QBERs with $x\ dB$ and $y\ dB$ attenuation should be less than the threshold. We get

\begin{equation}
\begin{aligned}
\{e_{x}, e_{y}\}< e_{th}.
\end{aligned}
\label{eq:exeyno}
\end{equation}

In detector control attack without blinding light, $P_{f,x}$ is defined as the detection probability  with full optical power when the attenuation is $x\ dB$. $P_{f,y}$ is likewise defined   when the attenuation  is $y\ dB$; similarly, with half power, $P_{h,x}$ and $P_{h,y}$ are defined as the detection probabilities when the attenuation are $x\ dB$ and $y\ dB$, respectively.  Suppose Eve select two measurement basis with equal probability. Then the detection rates with two attenuation values can be given by 

\begin{equation}
\begin{aligned}
R_{x}^{atk}=\frac{1}{4}P_{f,x}+\frac{1}{4}(2P_{h,x}),
\end{aligned}
\label{eq:Rmxattack}
\end{equation}

\begin{equation}
\begin{aligned}
R_{y}^{atk}=\frac{1}{4}P_{f,y}+\frac{1}{4}(2P_{h,y}).
\end{aligned}
\label{eq:Rmyattack}
\end{equation}

For simplicity, we analyse the case that both VA-SPDs have the same attenuation value ($x\ dB$ or $y\ dB$). Then the QBERs of Bob's one VA-SPD with $x\ dB$ and $y\ dB$ attenuation values are given by

\begin{equation}
\begin{aligned}
e_{x(s)}^{atk}=\frac{2P_{h,x}-P_{h,x}^{2}}{2P_{f,x}+2(2P_{h,x}-P_{h,x}^{2})},
\end{aligned}
\label{eq:eux}
\end{equation}

\begin{equation}
\begin{aligned}
e_{y(s)}^{atk}=\frac{2P_{h,y}-P_{h,y}^{2}}{2P_{f,y}+2(2P_{h,y}-P_{h,y}^{2})}.
\end{aligned}
\label{eq:euy}
\end{equation}
By substituting Eqs. \eqref{eq:Rmxattack}--\eqref{eq:Rmyattack} into  the Eqs. \eqref{eq:eux}--\eqref{eq:euy} respectively,  we get
\begin{equation}
\begin{aligned}
2P_{h,x}-P_{h,x}^{2}<2 \alpha^{\ast} e_{th}(P_{f,y}+2P_{h,y})-2e_{th}P_{h,x}^{2},
\end{aligned}
\label{eq:eux<=eth}
\end{equation}

\begin{equation}
\begin{aligned}
\alpha^{\ast}(2P_{h,y}-P_{h,y}^{2})<2\alpha^{\ast} e_{th}(P_{f,y}+2P_{h,y}-P_{h,y}^{2}).
\end{aligned}
\label{eq:euy<=eth}
\end{equation}
By adding both sides of these inequalities,  we  deduce that

\begin{widetext}
\begin{equation}
\begin{aligned}
(2P_{h,x}-P_{h,x}^{2})-4\alpha^{\ast}e_{th}P_{f,y}+\alpha^{\ast}(2-P_{h,y}-8e_{th})P_{h,y}+2e_{th}P_{h,x}^{2}+2\alpha^{\ast}e_{th}P_{h,y}^{2}< 0.
\end{aligned}
\label{eq:eux+euy}
\end{equation}
\end{widetext}
	As $ e_{th}<11\%$, $0\leq \{P_{f,x},\ P_{h,x},\ P_{f,y},\ P_{h,y} \}\leq 1 $ and $\alpha ^{\ast}> 1$,  we know that $(2P_{h,x}-P_{h,x}^{2})+\alpha^{\ast}(2-P_{h,y}-8e_{th})P_{h,y}+2e_{th}P_{h,x}^{2}+2\alpha^{\ast}e_{th}P_{h,y}^{2}\geq 0$, $-4\alpha^{\ast}e_{th}P_{f,y}\leq 0$. To make an effective  countermeasure criteria, It should be guaranteed that Eq. \eqref{eq:eux+euy} can not be satisfied for all the  values of $\alpha^{\ast}$, there are two following cases:

    	If $P_{h,x}\neq P_{f,y}$,  whether the Eq. \eqref{eq:eux+euy} can  be satisfied depends on the value of $P_{h,x}$, $P_{f,y}$, $\alpha^{\ast}$ and $P_{h,y}$, which means that the countermeasure criteria  is  not general.

  	If $P_{h,x}=P_{f,y}$,  then $(2P_{h,x}-P_{h,x}^{2})-4\alpha^{\ast}e_{th}P_{f,y}\geq 0$,  all the factors on the left  of Eq. \eqref{eq:eux+euy} is greater than 0, which is contradictory to the right result of Eq. \eqref{eq:eux+euy}. It means the two sub-cases:
 The one is the optical power before entering SPDs are equal,    then  half power with $x \ dB$ is equal to the full  power with $ y \ dB$, so  the difference of VA's value between $y$ and $x$ is $3\ dB$.  It meets the requirement of generalization of  criteria;
	The other one is the optical power before entering SPDs are different,  but their detection probabilities are  equal.  Therefore,  the countermeasure is influenced by the specific detector probabilities and is not general.

\section*{Appendix B: calculation process if one relationship is satisfied}
\label{AppendixB}
When both VA-SPDs in the same basis have the same attenuation value, and in the case that both QBERs ($e_{0(s)}^{atk}$ and $e_{3(s)}^{atk}$) are less than $ e_{th} $ (Eq. \eqref{eq:e0e3no} is satisfied), in order to deduce the range of $\frac{R_{0(s)}^{atk}}{R_{3(s)}^{atk}}$, let $\frac{1}{2e_{0(s)}^{atk}}-1=m_{1}$, $\frac{1}{2e_{3(s)}^{atk}}-1=m_{2}$. Then Eqs. \eqref{eq:e0attack} and \eqref{eq:e3attack} can be converted into

\begin{equation}
\begin{aligned}
P_{f,0}=m_{1}(2P_{h,0}-P_{h,0}^{2}),
\end{aligned}
\label{eq:Pfm1}
\end{equation}

\begin{equation}
\begin{aligned}
P_{f,3}=m_{2}(2P_{h,3}-P_{h,3}^{2}).
\end{aligned}
\label{eq:Phm2}
\end{equation}
With $0 \leq{P_{f,0}}\leq1 $ and $P_{h,0}=P_{f,3} $ we get
\begin{equation}
\begin{aligned}
0\leq{P_{h,0}}\leq1-\sqrt{1-\frac{1}{m_{1}}}.
\end{aligned}
\label{eq:BPhrange}
\end{equation}
Then $\frac{R_{0(s)}^{atk}}{R_{3(s)}^{atk}}$ is given by
\begin{equation}
\begin{aligned}
\frac{R_{0(s)}^{atk}}{R_{3(s)}^{atk}}=\frac{m_{1}(2P_{h,0}-P_{h,0}^{2})+2P_{h,0}}{P_{h,0}+2-2\sqrt{1-\dfrac{P_{h,0}}{m_{2}}}}.
\end{aligned}
\label{eq:BR0/2R3}
\end{equation}
Define $ m=\mathrm{min} \{m_{1}, m_{2}\} $,  then $m \geq\frac{1}{2e_{th}}-1$. Let $1-\sqrt{1-\frac{1}{m}}=x$, with Eq. \eqref{eq:BPhrange} we have
\begin{equation}
\begin{aligned}
\frac{R_{0(s)}^{atk}}{R_{3(s)}^{atk}}\geq \dfrac{m(2x-x^{2})+2x}{x+2-2\sqrt{1-\dfrac{x}{m}}}.
\end{aligned}
\label{eq:BR0/2R3range}
\end{equation}
We can simulate the lower bound of $\frac{R_{0(s)}^{atk}}{R_{3(s)}^{atk}}$, the result is shown with the red  line in Fig. \ref{Fig:euvsratio}.

Similarly, when both VA-SPDs in the same basis have the opposite attenuation value, and both  QBERs ($e_{0(opp)}^{atk}$ and $ e_{3(opp)}^{atk}$) are less than $e_{th}$, let $\frac{1}{2e_{0(opp)}^{atk}}-1=m_{3}$, $\frac{1}{2e_{3(opp)}^{atk}}-1=m_{4}$. Then Eqs. \eqref{eq:e0attack2} and \eqref{eq:e3attack2} can be converted into

\begin{equation}
\begin{aligned}
P_{f,0}=m_{3}(2P_{h,0}-P_{h,0}P_{h,3}),
\end{aligned}
\label{eq:Pfm1}
\end{equation}

\begin{equation}
\begin{aligned}
P_{f,3}=m_{4}(2P_{h,3}-P_{h,0}P_{h,3}).
\end{aligned}
\label{eq:Phm2}
\end{equation}
With Eq. \eqref{eq:Phm2}, we get $P_{h,0}=\frac{2m_{4}P_{h,3}}{1+m_{4}P_{h,3}}$, substitute it  in Eq. \eqref{eq:Pfm1}, as $0 \leq P_{f,0}\leq 1$ , we have

\begin{equation}
\begin{aligned}
0\leq \frac{2P_{h,3}-P_{h,3}^{2}}{1+m_{4}P_{h,3}}\leq \frac{1}{2m_{3}m_{4}}.
\end{aligned}
\label{eq:Pfm1range}
\end{equation}
Then we get the range of $P_{h,3}$
\begin{equation}
\begin{aligned}
0\leq P_{h,3} \leq \frac{4m_{3}m_{4}-m_{4}-\sqrt{(4m_{3}m_{4}-m_{4})^{2}-8m_{3}m_{4}}}{4m_{3}m_{4}}.
\end{aligned}
\label{eq:Ph3m1range}
\end{equation}
Then $\frac{R_{0(opp)}^{atk}}{R_{3(opp)}^{ atk}}$ is given by
\begin{equation}
\begin{aligned}
\frac{R_{0(opp)}^{atk}}{R_{3(opp)}^{ atk}}=\frac{m_{3}m_{4}(2-P_{h,3})+2m_{4}}{m_{4}+1+m_{4}P_{h,3}}.
\end{aligned}
\label{eq:R0R32range}
\end{equation}
Define $ m=\mathrm{min} \{m_{3}, m_{4}\} $, then $m \geq\frac{1}{2e_{th}}-1$, we have

\begin{equation}
\begin{aligned}
\frac{R_{0(opp)}^{atk}}{R_{3(opp)}^{ atk}}\geq \dfrac{4m^{2}+9m+m\sqrt{16m^{2}-8m-7}}{8m+3-\sqrt{16m^{2}-8m-7}}.
\end{aligned}
\label{eq:R0R322range}
\end{equation}
We simulate the lower bound of $\frac{R_{0(opp)}^{atk}}{R_{3(opp)}^{atk}}$, the result is shown with the blue dashed line in Fig. \ref{Fig:euvsratio}.

Under the detector control attack, since  Eve could control transmittance and number of trigger pulses to guarantee the detection rates unchanged, $t$ is the attack transmission parameter which satisfies $t \geq 1$,  then we have

\begin{equation}
\begin{aligned}
R_{0}^{atk}=tR_{0}.
\end{aligned}
\label{eq:R0m=tR0}
\end{equation}

In the case that the relationship of Eq. \ref{eq:R0R3no} is satisfied,  \eqref{eq:R0attack} and \eqref{eq:R3attack} can be converted into

\begin{equation}
\begin{aligned}
P_{f,0}+2P_{h,0}=4tR_{0},
\end{aligned}
\label{eq:CPfhR0}
\end{equation}

\begin{equation}
\begin{aligned}
P_{f,3}+2P_{h,3}=\frac{4tR_{0}}{\alpha}.
\end{aligned}
\label{eq:CPhqR0}
\end{equation}
As $ 0 \leq $ \{$P_{f,0}$, $P_{h,0}$,  $P_{f,3}$, $P_{h,3}$\} $ \leq 1$, by using Eqs. \eqref{eq:CPfhR0} and \eqref{eq:CPhqR0}, we have
\begin{equation}
\begin{aligned}
0\leq tR_{0}\leq 0.75.
\end{aligned}
\label{eq:therangetR0}
\end{equation}

When both VA-SPDs in the same basis have the same attenuation value, then  $e_{0(s)}^{atk}$, $e_{3(s)}^{atk}$ can be converted into
\begin{equation}
\begin{aligned}
e_{0(s)}^{atk}=\frac{2P_{h,0}-P_{h,0}^{2}}{8tR_{0}-2P_{h,0}^{2}},
\end{aligned}
\label{eq:Ceu0PHR0}
\end{equation}

\begin{equation}
\begin{aligned}
e_{3(s)}^{atk}=\frac{2P_{h,3}-P_{h,3}^{2}}{\dfrac{8tR_{0}}{\alpha}-2P_{h,3}^{2}}.
\end{aligned}
\label{eq:Ceu3PHR0}
\end{equation}
According  Eq. \eqref{eq:Ceu3PHR0} and Eq. \eqref{eq:CPhqR0}, we  get

\begin{widetext}
\begin{equation}
\begin{aligned}
P_{f,3}=P_{h,0}=\frac{1+\dfrac{4tR_{0}}{\alpha}e_{3(s)}^{atk}-\dfrac{2tR_{0}}{\alpha}-\sqrt{(1+\dfrac{4tR_{0}}{\alpha}e_{3(s)}^{atk}-\dfrac{2tR_{0}}{\alpha})^{2}+2(e_{3(s)}^{atk}-\dfrac{1}{2})^{2}(\dfrac{8tR_{0}}{\alpha}-8(\dfrac{tR_{0}}{\alpha})^{2})}}{e_{3(s)}^{atk}-\dfrac{1}{2}},
\end{aligned}
\label{eq:CPhrange}
\end{equation}
\end{widetext}
thus we substitute  Eq. \eqref{eq:CPhrange} to
Eq. \eqref{eq:Ceu0PHR0}. We can simulate the relationship of the QBERs with  $0 \ dB$ ($e_{0(s)}^{atk}$) and $3\ dB$ ($e_{3(s)}^{atk}$), we set $tR_{0}=0.75$ for Eq. \eqref{eq:therangetR0}, because $e_{0(s)}^{atk}$ and $e_{3(s)}^{atk}$ are increasing with decreasing $tR_{0}$. If Eq. \eqref{eq:CPhrange} is larger than 1 (smaller than 0), we take 1(0) for $P_{f,3}$. The result is shown with red and blue  lines in Fig. \ref{Fig:e0vse3}.

Similarly, when both VA-SPDs in the same basis have the  opposite attenuation value,  $e_{0(opp)}^{atk}$, $e_{3(opp)}^{atk}$ can be converted into
\begin{equation}
\begin{aligned}
e_{0(opp)}^{atk}=\frac{2P_{h,0}-P_{h,0}P_{h,3}}{8tR_{0}-2P_{h,0}P_{h,3}},
\end{aligned}
\label{eq:Ceu0PHR022}
\end{equation}

\begin{equation}
\begin{aligned}
e_{3(opp)}^{atk}=\frac{2P_{h,3}-P_{h,0}P_{h,3}}{\dfrac{8tR_{0}}{\alpha}-2P_{h,0}P_{h,3}}.
\end{aligned}
\label{eq:Ceu3PHR02}
\end{equation}
According Eqs. \eqref{eq:CPhqR0} and \eqref{eq:Ceu3PHR02}, we get
 $P_{h,3}=\dfrac{2tR_{0}}{\alpha}-\dfrac{1}{2}P_{f,3}$.
 \begin{widetext}
\begin{equation}
\begin{aligned}
 P_{f,3}=P_{h,0}=\frac{\dfrac{2tR_{0}}{\alpha}+1-\dfrac{4tR_{0}}{\alpha}e_{3(opp)}^{atk}-\sqrt{(\dfrac{4tR_{0}}{\alpha}e_{3(opp)}^{atk}-\dfrac{2tR_{0}}{\alpha}-1)^{2}-4(e_{3(opp)}^{atk}-\dfrac{1}{2})^{2}(\dfrac{8tR_{0}}{\alpha})}}{1-2e_{3(opp)}^{atk}}.
 \end{aligned}
\end{equation}  
 \end{widetext}
Take these equation to Eq. \eqref{eq:Ceu0PHR022}. We can simulate the relationship of the QBERs with  $0 \ dB$ ($ e_{0(opp)}^{atk} $) and $3\ dB$ ($ e_{3(opp)}^{atk} $), we set $tR_{0}=0.75$ for Eq. \eqref{eq:therangetR0}, because $e_{0(opp)}^{atk}$ and $e_{3(opp)}^{atk}$ are increasing with decreasing $tR_{0}$. Similarly, if Eq. \eqref{eq:CPhrange} is larger than 1(smaller than 0), we take 1(0) for $P_{f,3}$. The result is shown with red and blue dashed lines in Fig. \ref{Fig:e0vse3}.

\nocite{*}


\end{document}